\documentclass[conference]{IEEEtran}
\IEEEoverridecommandlockouts
\usepackage{cite}
\usepackage{amsmath,amssymb,amsfonts}
\usepackage{algorithmic}
\usepackage{graphicx}
\usepackage{textcomp}
\usepackage{xcolor}
\usepackage{stfloats}
\usepackage{url}
\usepackage{verbatim} 
\usepackage{booktabs}
\usepackage{cite}
\usepackage{tabularx}
\usepackage{multirow}
\def\BibTeX{{\rm B\kern-.05em{\sc i\kern-.025em b}\kern-.08em
    T\kern-.1667em\lower.7ex\hbox{E}\kern-.125emX}}

\begin{document}

\title{Data-Efficient Motor Condition Monitoring with Time Series Foundation Models\\}

\author{Deyu Li\textsuperscript{*}, Xinyuan Liao\textsuperscript{†}, Shaowei Chen\textsuperscript{*}, and Shuai Zhao\textsuperscript{‡}\\
Email: deyuli@mail.nwpu.edu.cn, cgong@nwpu.edu.cn, xin-yuan.liao@connect.polyu.hk, szh@energy.aau.dk\\
\textsuperscript{*}Electronics and Information, Northwestern Polytechnical University, Xi'an, China,\\
\textsuperscript{†}Electrical and Electronic Engineering, The Hong Kong Polytechnic University, Kowloon, Hong Kong,\\
\textsuperscript{‡}AAU Energy, Aalborg University, Aalborg 9220, Denmark.
\thanks{This paper has been accepted for IEEE APEC 2026.}
}

\maketitle

\begin{abstract}
Motor condition monitoring is essential for ensuring system reliability and preventing catastrophic failures. However, data-driven diagnostic methods often suffer from sparse fault labels and severe class imbalance, which limit their effectiveness in real-world applications. This paper proposes a motor condition monitoring framework that leverages the general features learned during pre-training of two time series foundation models, MOMENT and Mantis, to address these challenges. By transferring broad temporal representations from large-scale pre-training, the proposed approach significantly reduces dependence on labeled data while maintaining high diagnostic accuracy. Experimental results show that MOMENT achieves nearly twice the performance of conventional deep learning models using only 1\% of the training data, whereas Mantis surpasses state-of-the-art baselines by 22\%, reaching 90\% accuracy with the same data ratio. These results demonstrate the strong generalization and data efficiency of time series foundation models in fault diagnosis, providing new insights into scalable and adaptive frameworks for intelligent motor condition monitoring.
\end{abstract}

\begin{IEEEkeywords}
Power Electronics, Condition Monitoring, Artificial Intelligence, Time Series Foundation Models.
\end{IEEEkeywords}

\section{Introduction}
\IEEEPARstart{R}{eliable} health monitoring is essential for ensuring the safety, efficiency, and longevity of electric drive systems. Electric motors—particularly permanent magnet synchronous motors (PMSMs)—serve as the core components in electric vehicles, industrial drives, renewable energy systems, and high-speed elevators \cite{luo2019model}. Operating under fluctuating loads, thermal stress, and electromagnetic disturbances, these motors are prone to mechanical wear, insulation degradation, and imbalance faults \cite{yoon2016high}. Without timely detection, such degradations can escalate into catastrophic failures, leading to costly downtime and system damage. Consequently, accurate and real-time fault diagnosis has become a critical requirement for achieving predictive maintenance and high system reliability.

Over the past decade, advances in deep learning and AI have made data-driven condition monitoring a rising focus in power electronics~\cite{zhao2020overview}. Models such as convolutional and recurrent neural networks have been successfully applied to inverter and motor fault diagnosis. For example, Cai et al.~\cite{cai2017} employed Bayesian networks for inverter fault diagnosis, and Zhang et al.~\cite{zhang2021} proposed a few-shot federated learning method for PMSM fault detection. However, these models are typically task-specific, requiring large volumes of labeled data and retraining for each motor type, load condition, or fault category. This lack of generalization severely limits their scalability and adaptability in practical industrial scenarios.

To overcome these limitations, the generalization about foundation models has recently emerged as a promising solution. Large-scale pre-trained models such as BERT~\cite{kenton2019bert} and GPT~\cite{achiam2023gpt} have demonstrated strong generalization and transfer capabilities across domains, enabling rapid adaptation to new tasks with limited supervision. Extending this concept to temporal domains, time series foundation models (TSFMs)~\cite{Yuqietal-2023-PatchTST} achieve similar flexibility through self-supervised learning. Representative examples of self-supervised learning methods include masked reconstruction (MOMENT~\cite{goswami2024moment}), autoregressive learning (Timer~\cite{liu2024timer}), and contrastive learning (Mantis~\cite{feofanov2025mantis}). Furthermore, using minimal (or few-shot) supervised fine-tuning strategies such as LoRA~\cite{hu2022lora} allows these models to adapt efficiently to downstream tasks like fault diagnosis~\cite{lin2025fd} and condition monitoring~\cite{liao2025petsfm}.

Building on these developments, this work explores the application of pre-trained TSFMs for motor condition monitoring. The proposed framework bridges general-purpose time-series modeling with domain-specific diagnostic tasks, enhancing adaptability and scalability. It establishes a systematic approach validated on real motor dataset, introducing a quantitative evaluation method to assess model generalization and guide the selection of foundation models, and demonstrates scalability through consistent performance gains with increasing data volume and model capacity. The code details accompanying the paper are open-sourced on GitHub \footnote{[Online]. https://github.com/ms140429/FM4Motor}.

The remainder of this paper is organized as follows:  Section~\ref{fault} details the experimental setup. Section~\ref{method} presents the proposed framework, including model assessment and fine-tuning strategies. Section~\ref{case} describes the results and discusses this framework. Section~\ref{cons} concludes this work.

\section{Experimental Setting for Motor Fault Diagnosis}\label{fault}
\subsection{The Occurrence of Stator Winding Faults}
\begin{figure}[!t]
    \centering
    \includegraphics[width=1.0\columnwidth]{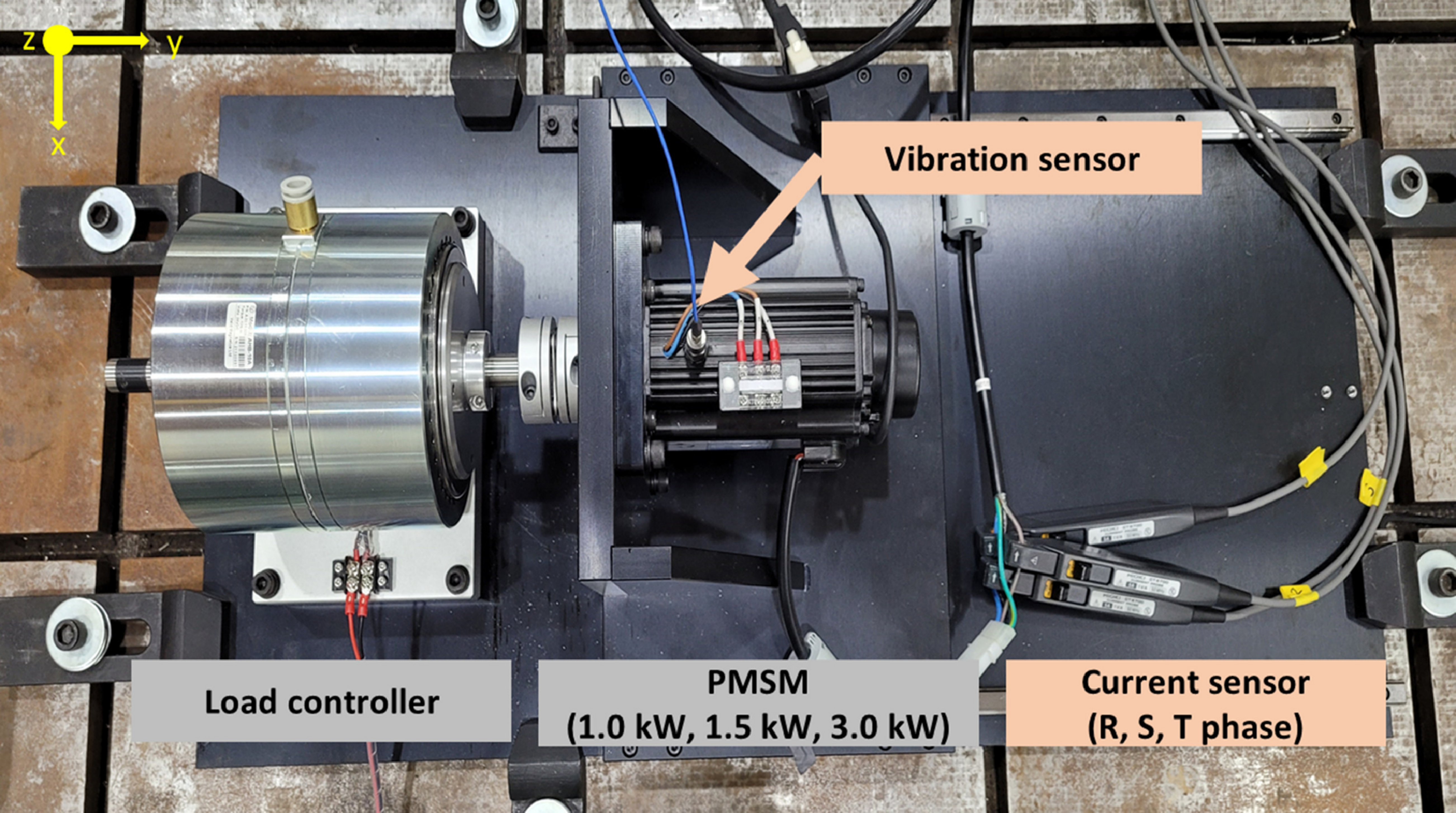}
    \caption{Experimental setup for PMSM fault data acquisition.}
    \label{fig:real}
    \vspace{-0.2cm}
\end{figure}

This study utilizes a publicly available dataset of PMSM fault signals~\cite{jung2023vibration}. The test platform records both vibration and current signals under healthy and faulty conditions across multiple load levels. As shown in Fig.~\ref{fig:real}, the experimental setup comprises a PMSM, a hysteresis brake, and various sensors. A Valid Magnetics AHB-10A hysteresis brake applies torque loads up to 10~Nm, while flexible couplings and linear guides are employed to prevent shaft misalignment between the load controller and the PMSM.

The primary fault type investigated in this work is partial stator winding failure, including inter-coil and inter-turn short-circuit faults. These faults introduce parasitic current paths that divert part of the drive current from the stator windings, leading to reduced electromagnetic field strength and diminished torque output according to Kirchhoff’s current law. Among common electrical faults, these are particularly challenging to detect due to their gradual onset and weak external manifestations.

\begin{figure}[!t]
    \centering
    \includegraphics[width=0.8\columnwidth]{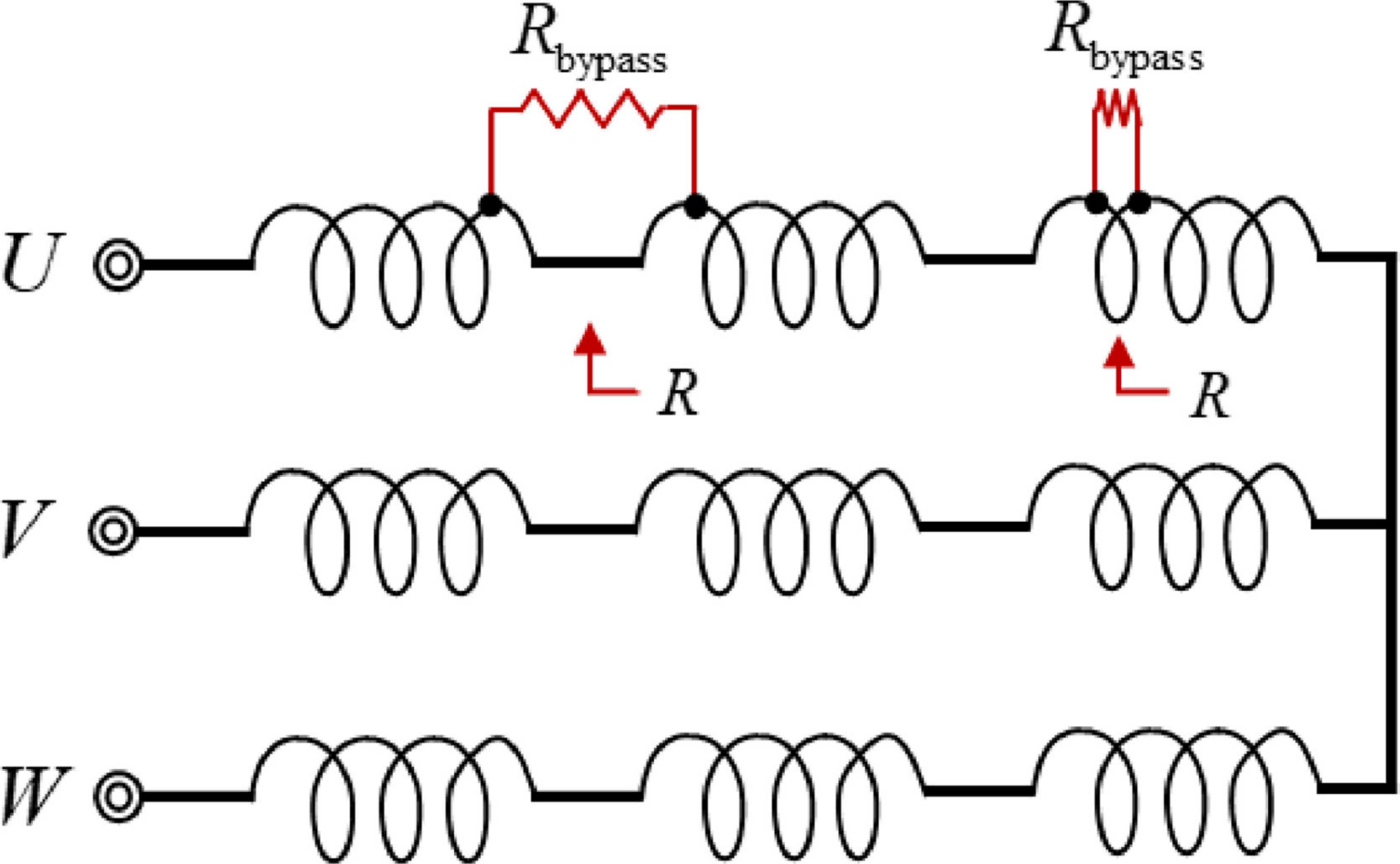}
    \caption{Stator winding of short circuit fault scheme where $R_\text{bypass}$ is bypassing resistance, and $R$ is stator circuit resistance.}
    \label{fig:r}
    \vspace{-0.2cm}
\end{figure}

To reproduce such faults experimentally, the test bench inserts controllable bypass resistors into the stator circuit. The lower the resistor value, the higher the fault severity, as more current flows through the bypass path. Specifically, inter-turn faults are created by connecting the bypass resistor between the first and second turns of the U-phase, while inter-coil faults connect it between the last and first turns of adjacent coils as shown in Fig.~\ref{fig:r}. The fault severity is expressed as:

\begin{equation}
\text{FaultRatio (FR)} = \left( \frac{R}{R + R_{\mathrm{bypass}}} \right) \times 100
\end{equation}
where $R_{\mathrm{bypass}}$ is the bypass resistance value and $R$ is the resistance value of the stator.

\subsection{Dataset Description and Preprocessing}
Building upon the fault mechanisms described in the previous section, the stator winding faults—inter-coil and inter-turn short circuits—were experimentally reproduced using a PMSM fault test bench \cite{jung2023vibration}. The dataset includes three PMSMs rated at 1.0~kW, 1.5~kW, and 3.0~kW, each tested under sixteen operating conditions encompassing healthy, inter-coil, and inter-turn fault states. Although two types of short-circuit fault were originally recorded, the first two measurements at each power level (corresponding to 0\% fault severity) were merged into the category of normal conditions. After verification and consolidation, a total of 45 valid classes were obtained from the original 48 recordings.

\begin{figure}[!t]
\centering
\includegraphics[width=1.0\columnwidth]{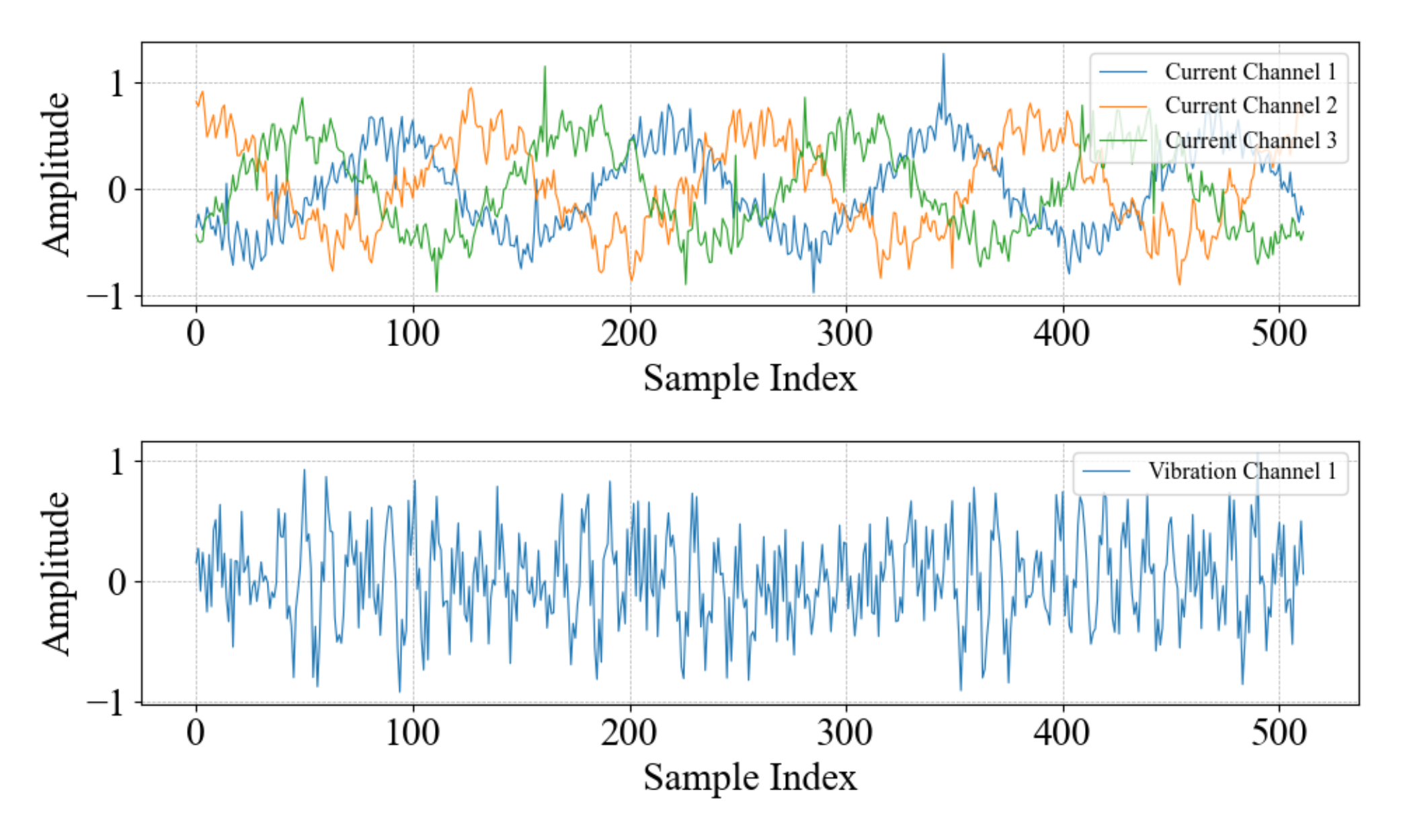}
\caption{Example of aligned PMSM fault signals: three-phase currents (Channels~1–3) and one vibration signal (Channel~4).}
\label{fig:input_signals}
\vspace{-0.2cm}
\end{figure}

The test bench simultaneously captured three-phase current and single-axis vibration signals, yielding approximately 576 million and 147 million data points, respectively. To ensure synchronization and model compatibility, the current signals within a sample are downsampled from 2000~Hz to 512~Hz to match the vibration sampling rate. Since TSFMs accept fixed-length inputs of 512 time steps, the aligned current and vibration signals were segmented and combined into four input channels. Fig.~\ref{fig:input_signals} illustrates an example of one aligned input sample, resulting in roughly 960,000 samples.

\begin{figure}[!t]
\centering
\includegraphics[width=1.0\columnwidth]{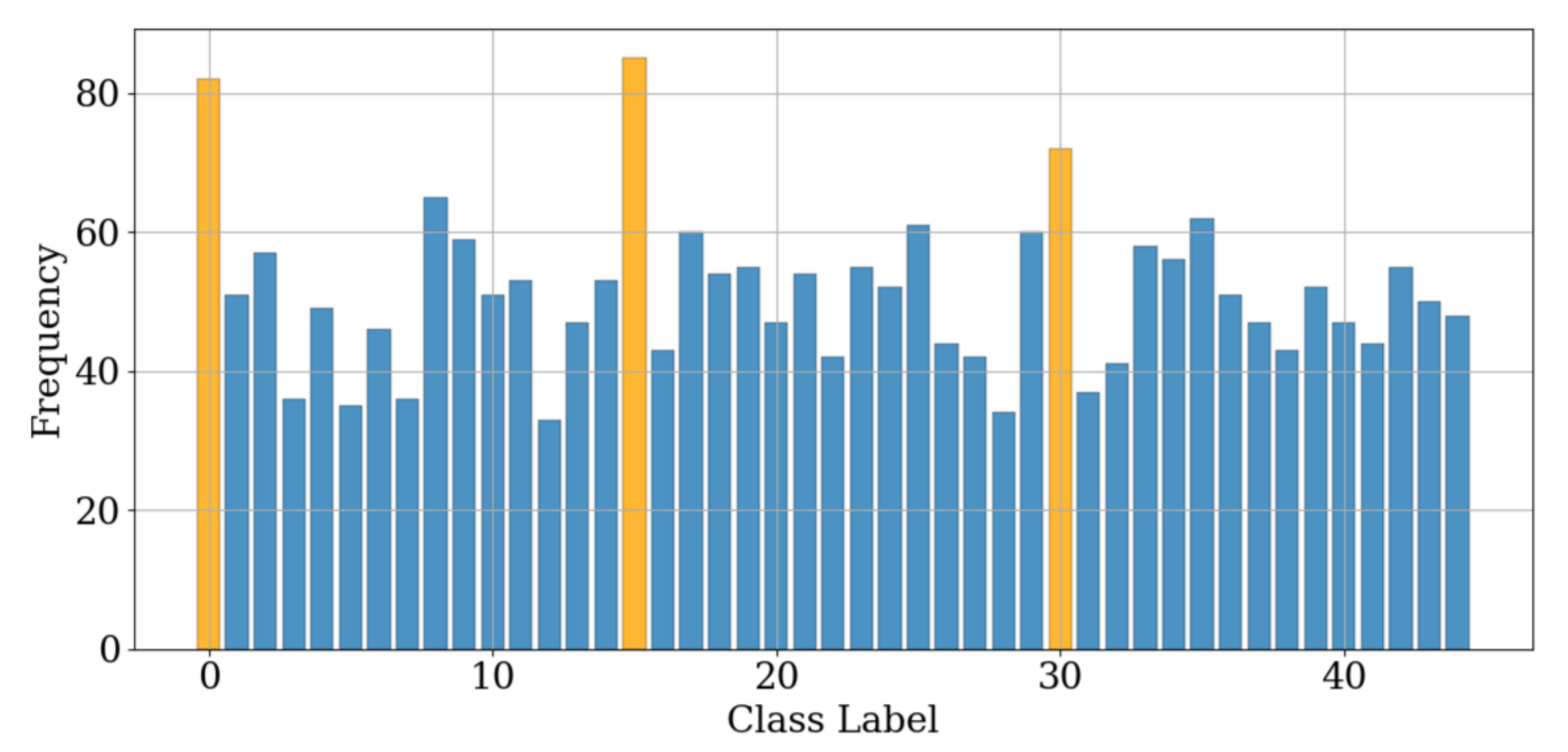}
\caption{Label distribution of the 1\% training subset used for model evaluation. Yellow denotes normal samples, while blue indicates fault samples with varying degrees of defect severity.}
\label{fig:label_hist}
\vspace{-0.2cm}
\end{figure}

The dataset was randomly partitioned into an 80\% training set and a 20\% test set.  To emulate real-world industrial scenarios characterized by scarce labeled data and class imbalance, training experiments primarily utilized 1\% and 5\% subsets of the training dataset. Fig.~\ref{fig:label_hist} shows the label distribution for the 1\% subset, where the normal category includes approximately 80 samples and each fault category around 45 samples.

\section{Proposed Motor Fault Diagnosis Framework} \label{method}
\subsection{Overview of the Process} \label{pre}
Time series foundation models (TSFMs) exhibit strong cross-domain generalization, making them promising for industrial diagnostics. However, as existing TSFMs are still under rapid development and less mature than large language models, a practical and lightweight workflow is required to identify suitable models for motor condition monitoring and perform efficient adaptation.

The proposed process consists of two main stages. In the first stage, a small subset of labeled samples from the target dataset is used to evaluate multiple pre-trained models within the foundation model repository. The LogME~\cite{you2021logme} metric is employed to quantify the compatibility between each pre-trained model and the target task, allowing the selection of the most transferable candidate. In the second stage, the selected model undergoes parameter-efficient~\cite{hu2022lora} or full fine-tuning~\cite{chen2024overview} to obtain diagnostic performance metrics such as classification accuracy.

\begin{figure}[!t]
    \centering
    \includegraphics[width=1.0\columnwidth]{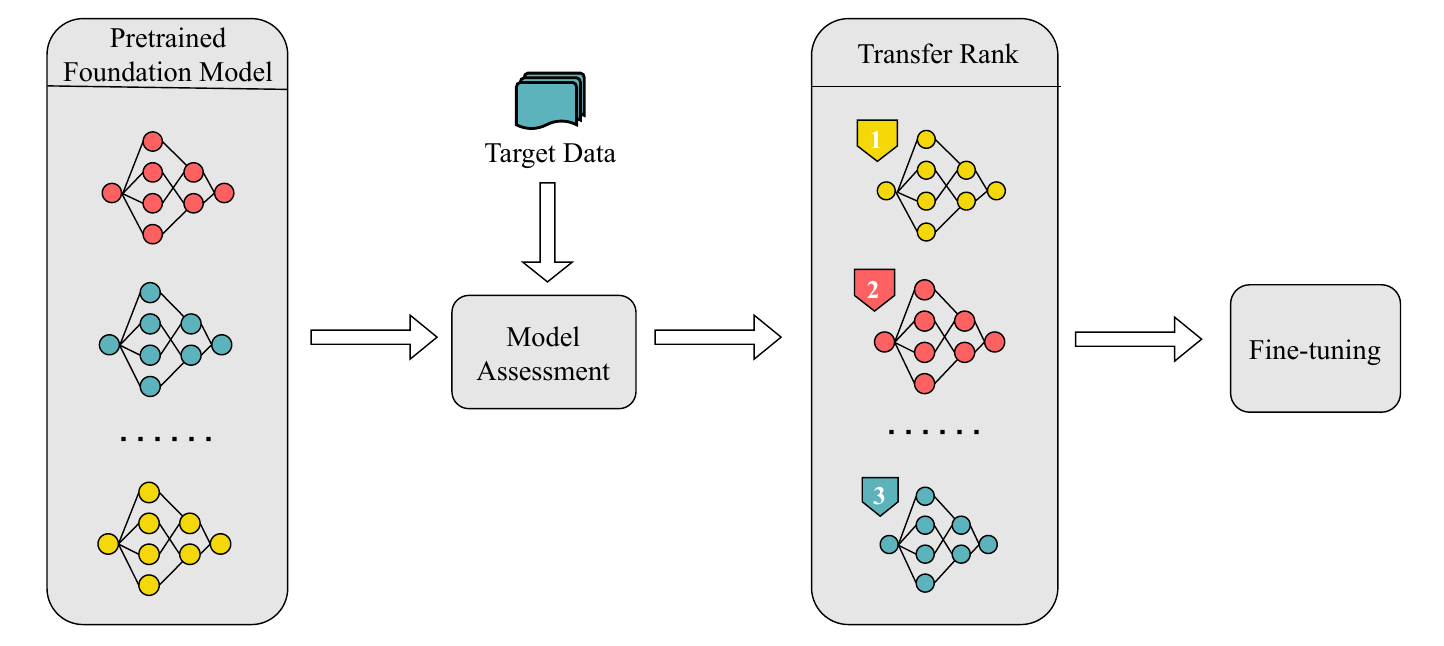} 
    \caption{Proposed workflow for adapting pre-trained TSFMs to motor fault diagnosis, enabling efficient model selection and data-efficient fine-tuning.}
    \label{fig:m1}
    \vspace{-0.2cm}
\end{figure}

As illustrated in Fig.~\ref{fig:m1}, this workflow enables a systematic and data-efficient pathway for applying TSFMs to motor fault diagnosis. It provides a quantitative measure of model generalization while avoiding exhaustive trial-and-error across numerous model configurations. The subsequent sections detail the fine-tuning process and experimental validation.

\subsection{Model Assessment}
\begin{figure}[!t]
    \centering
    \includegraphics[width=1.0\columnwidth]{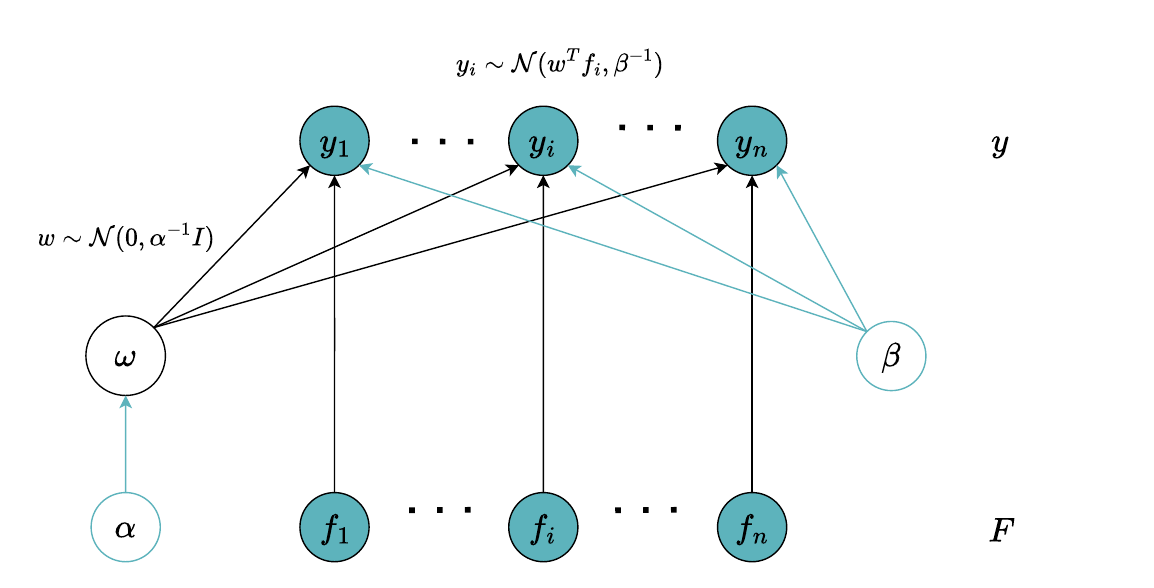}
    \caption{Illustration of the LogME model assessment process for foundation models in power electronics.}
    \label{fig:m3}
    \vspace{-0.2cm}
\end{figure}

Selecting an appropriate pre-trained model for a downstream task often requires detailed knowledge of model architectures and their pre-training objectives, a requirement that is impractical in industrial environments. To streamline this process, this work employs LogME~\cite{you2021logme}, an efficient metric for estimating the adaptability of pre-trained models without additional fine-tuning. Under this framework, each foundation model is treated purely as a feature extractor: a single forward pass through the available dataset yields the feature matrix and corresponding labels. LogME then evaluates the marginal likelihood linking these features to the label distribution, producing a scalar score that reflects the expected performance of the model on the target dataset. This enables fast and systematic model selection from large repositories of pre-trained temporal models. The overall evaluation pipeline is summarized in Fig.~\ref{fig:m3}.

Let $f_i$ denote the feature vector of the $i$-th sample, $y_i$ the corresponding label, and $\omega$ the model parameters. Assuming a Gaussian prior $P(\omega|\alpha) = N(\omega|0, \alpha^{-1} I)$ and Gaussian likelihood $P(y_i|f_i, \omega, \beta) = N(y_i|\omega^T f_i, \beta^{-1})$, the conditional probability can be expressed as

\begin{equation}
P(y|F, \omega, \beta) = \prod_{i=1}^n N(y_i|\omega^T f_i, \beta^{-1})
\end{equation}
where $F$ represents the feature matrix and $y$ the set of all labels. Integrating over $\omega$ yields

\begin{equation}
\begin{split}
p(y|F,\alpha,\beta) &= \int p(\omega|\alpha), p(y|F,\omega,\beta), d\omega \\
&= \left(\frac{\beta}{2\pi}\right)^{\frac{n}{2}} \left(\frac{\alpha}{2\pi}\right)^{\frac{D}{2}}
\int e^{-\frac{\alpha}{2}\omega^T\omega - \frac{\beta}{2}|F\omega - y|^2} d\omega
\end{split}
\end{equation}
where $D$ denotes the feature dimension. Taking the logarithm, the log marginal likelihood is obtained as

\begin{equation}
\begin{split}
\mathcal{L}(\alpha, \beta) &= \log p(y|F,\alpha,\beta) \\
&= \frac{n}{2}\log\beta + \frac{D}{2}\log\alpha - \frac{n}{2}\log 2\pi \\
&\quad - \frac{\beta}{2}|F m - y|^2 - \frac{\alpha}{2} m^T m - \frac{1}{2}\log|A|
\end{split}
\end{equation}
where $m = \beta A^{-1} F^T y$ and $A = \beta F^T F + \alpha I$. The LogME value corresponds to $\mathcal{L}(\alpha, \beta)$ after maximizing with respect to $\alpha$ and $\beta$. Through the use of singular value decomposition (SVD), the computational cost is further reduced, enabling efficient evaluation even for large-scale model repositories.

\subsection{Supervised Fine-Tuning}

\begin{table*}[!t]
\centering
\caption{Comparison of Parameter-Efficient and Full Fine-Tuning Strategies}
\label{tab:finetuning_comparison}
\footnotesize
\begin{tabularx}{\textwidth}{@{}lXX@{}}
\toprule
\textbf{Aspect} & \textbf{Parameter-Efficient Fine-Tuning (LoRA)} & \textbf{Full Fine-Tuning} \\
\midrule
\textbf{Mechanism} & Freezes $W_0$ and trains low-rank matrices $(A,B)$ for task-specific adaptation. & Unfreezes and optimizes all parameters for full-domain adaptation. \\
\textbf{Trainable Parameters} & $r(d+k)$ (low) & $d \times k$ (high) \\
\textbf{Resource Demand} & Low memory and computation overhead. & High computational and memory cost. \\
\textbf{Adaptation Performance} & Efficient with limited labeled data; strong transferability. & Superior performance when data and compute are abundant. \\
\textbf{Deployment Efficiency} & Adapted weights can be merged; no added inference latency. & Requires larger storage and inference cost. \\
\bottomrule
\end{tabularx}
\end{table*}

After selecting the optimal foundation model, fine-tuning is performed to adapt the general representation to task-specific motor fault patterns. Following the approaches in~\cite{chen2024overview}, two strategies are considered: parameter-efficient fine-tuning and full fine-tuning.

\begin{figure}[!t]
    \centering
    \includegraphics[width=1.0\columnwidth]{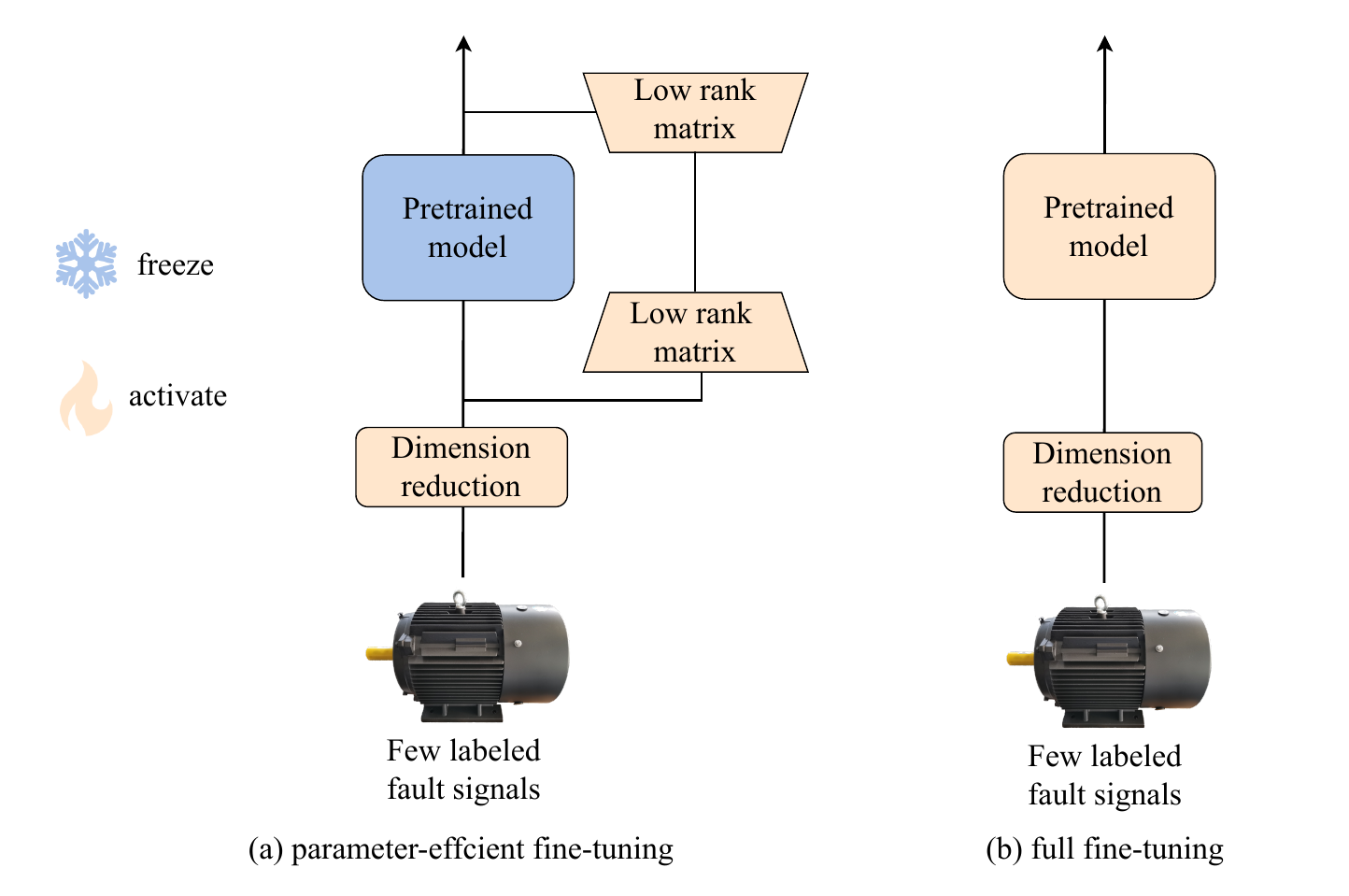}
    \caption{Fine-tuning strategies for foundation models in motor fault diagnosis.}
    \label{fig:m2}
    \vspace{-0.2cm}
\end{figure}

\subsubsection{Parameter-Efficient Fine-Tuning}
To mitigate the prohibitive cost of training large models, this work adopts the Low-Rank Adaptation (LoRA) technique, which freezes the pre-trained weights $W_0$ and injects trainable low-rank matrices to capture task-dependent knowledge. The adapted weight is expressed as
\begin{equation}
    W = W_0 + B A
\end{equation}
where $A \in \mathbb{R}^{r \times k}$ and $B \in \mathbb{R}^{d \times r}$ are trainable matrices and $r \ll \min(d,k)$ is the predefined rank. For input $x$, the output becomes
\begin{equation}
    y = W x = W_0 x + B A x
\end{equation}

Only $r(d+k)$ parameters are updated, yielding substantial reductions in memory and computation. Moreover, $W_0 + B A$ can be merged for inference without additional latency, preserving deployment efficiency.

\subsubsection{Full Fine-Tuning}
Full fine-tuning unfreezes and optimizes all parameters of the pre-trained backbone. While computationally intensive, this strategy provides maximum flexibility, allowing the model to fully adapt to target-domain distributions. It typically achieves superior performance when sufficient data and resources are available. The general fine-tuning framework, which integrates both LoRA-based and full-finetuning, is illustrated in Table~\ref{tab:finetuning_comparison} and  Fig.~\ref{fig:m2}.

\section{Results and Discussion} \label{case}
\subsection{Selection of Foundation Models}
To examine the generalization performance of different TSFMs within the proposed framework, two representative architectures are selected: MOMENT~\cite{goswami2024moment} and Mantis~\cite{feofanov2025mantis}. MOMENT is based on masked reconstruction, a self-supervised pre-training strategy that leverages large-scale datasets and extensive model parameters to enhance multi-task adaptability, including classification, prediction, interpolation, and anomaly detection. In contrast, Mantis employs contrastive learning on a smaller corpus with fewer parameters, prioritizing discriminative feature extraction for classification-oriented tasks.

\begin{figure}[!t]
    \centering
    \includegraphics[width=1.0\columnwidth]{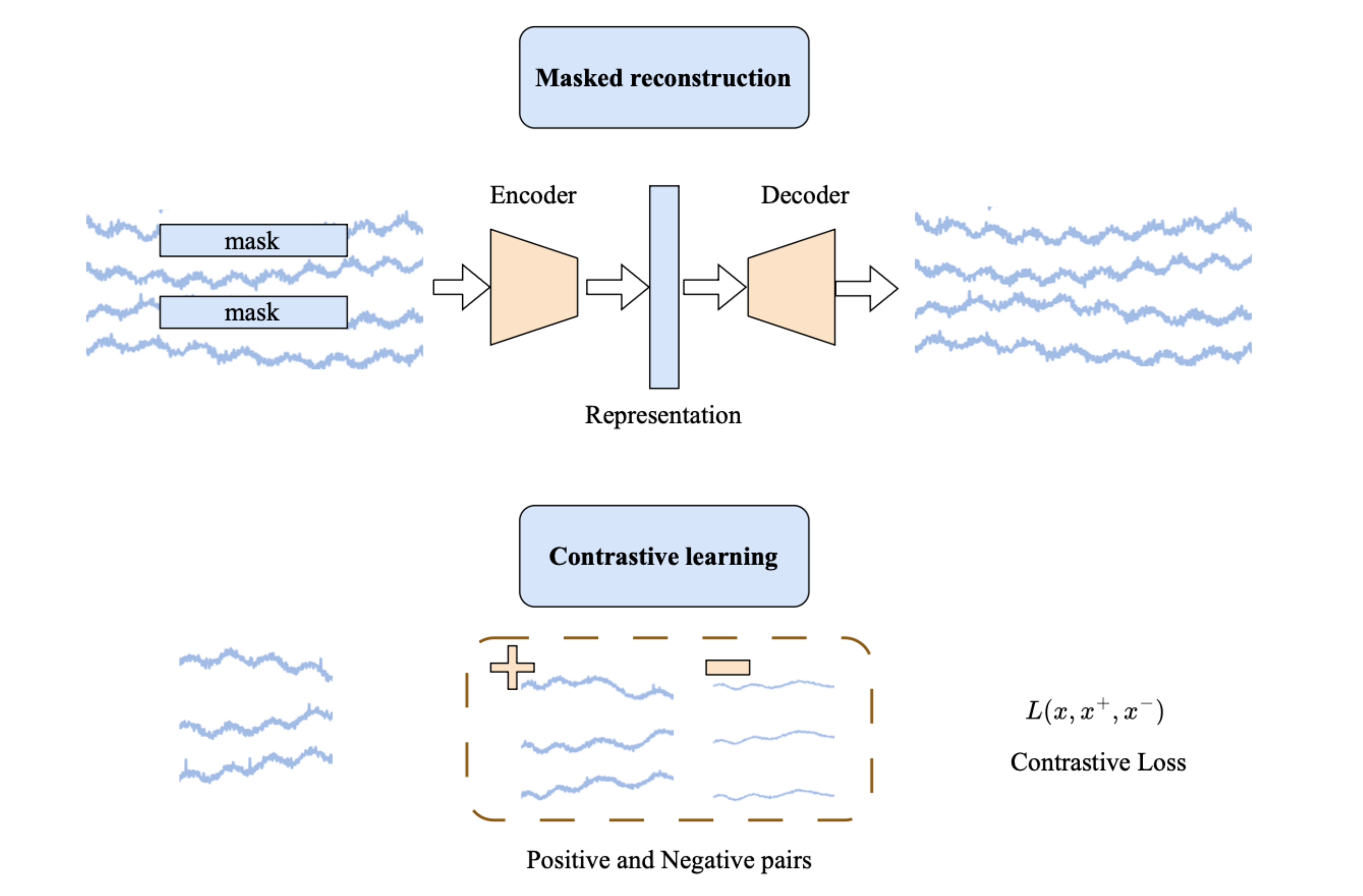}
    \caption{Overview of the self-supervised learning framework.}
    \label{fig}
    \vspace{-0.2cm}
\end{figure}

These two models embody distinct self-supervised paradigms and differ significantly in network scale and data dependency. Their inclusion enables a balanced evaluation of both general-purpose and task-specific pre-training approaches. This comparison thus serves two objectives: (1) to validate the effectiveness of the proposed assessment framework, and (2) to provide empirical guidelines for selecting suitable foundation models in motor fault diagnosis and related industrial applications. The overall self supervised learning methods are summarized in Fig.~\ref{fig}.

\subsection{Model Assessment}
LogME is adopted to quantitatively assess the transferability of the MOMENT and Mantis models by estimating the marginal likelihood of labeled samples. Table~\ref{tab:model_performance} reports the LogME scores computed using progressively increased data ratios. The relative improvement rate is calculated as the ratio of the score difference between the two models to the lower score, providing a clear measure of performance gain. Mantis consistently achieves higher scores across all settings, indicating stronger adaptability to the downstream fault classification task and confirming the advantage of contrastive pre-training in extracting discriminative features.

\begin{table}[!t]
    \centering
    \caption{LogME-Based Assessment Scores of MOMENT and Mantis Models}
    \begin{tabular}{l*{5}{c}}
        \toprule
        Data Ratio & 5\% & 10\% & 20\% & 50\% & 80\% \\
        \midrule
        MOMENT & 0.6274 & 0.6459 & 0.6666 & 0.6881 & 0.6973 \\
        Mantis & 0.8765\textsuperscript{*} & 0.9086\textsuperscript{*} & 0.9322\textsuperscript{*} & 0.9492\textsuperscript{*} & 0.9534\textsuperscript{*} \\
        \bottomrule
        \multicolumn{6}{l}{\footnotesize *Improvement: 39.7\%, 40.7\%, 39.9\%, 38.0\%, and 36.7\%, respectively.} \\
    \end{tabular}
    \label{tab:model_performance}
\end{table}

To further validate representational effectiveness, features from each model are evaluated using multiple classical classifiers, including Support Vector Machines (SVM), Random Forests (RF), Decision Trees (DT), and $k$-Nearest Neighbors (k-NN), with grid-searched hyperparameters. Baseline performance is obtained by training SVM directly on raw data. The classification results in Table~\ref{tab:test_accuracy} demonstrate substantial improvements from both foundation models compared to raw signals. For example, at the 1\% data ratio, SVM accuracy improves from 39.7\% (raw) to 40.5\% with MOMENT and 84.1\% with Mantis; at 5\% data, Mantis further increases accuracy to 90.9\%.

\begin{table}[!t]
    \centering
    \caption{Classification Accuracy (\%) of Different Classifiers Using MOMENT and Mantis Feature Extraction}
    \label{tab:test_accuracy}
    \begin{tabular}{@{}lccccc@{}}
    \toprule
    \multicolumn{6}{c}{\textbf{MOMENT}} \\
    \cmidrule(l){1-6}
    Data Ratio & Raw in SVM & k-NN & DT & RF & \textbf{SVM} \\ 
    \midrule
    1\% & 39.7 & 24.4 & 20.0 & 36.5 & \textbf{40.5} \\ 
    5\% & 52.7 & 31.6 & 27.9 & 43.9 & \textbf{53.9} \\ 
    10\% & 52.8 & 31.8 & 28.2 & 44.3 & \textbf{54.5} \\ 
    20\% & 52.8 & 31.4 & 27.9 & 44.3 & \textbf{54.6} \\ 
    50\% & 52.4 & 31.5 & 27.8 & 44.4 & \textbf{54.5} \\ 
    80\% & 52.7 & 31.4 & 27.1 & 44.4 & \textbf{54.4} \\ 
    \bottomrule
    \end{tabular}
    \quad
    \begin{tabular}{@{}lccccc@{}}
    \toprule
    \multicolumn{6}{c}{\textbf{Mantis}} \\
    \cmidrule(l){1-6} 
    Data Ratio & Raw in SVM & k-NN & DT & RF & \textbf{SVM} \\ 
    \midrule
    1\% & 39.7 & 62.1 & 56.2 & 73.8 & \textbf{84.1} \\ 
    5\% & 52.7 & 70.0 & 64.5 & 80.7 & \textbf{90.9} \\ 
    10\% & 52.8 & 70.4 & 64.7 & 80.4 & \textbf{91.0} \\ 
    20\% & 52.8 & 70.3 & 65.1 & 80.4 & \textbf{91.0} \\ 
    50\% & 52.4 & 69.8 & 64.8 & 80.4 & \textbf{90.8} \\ 
    80\% & 52.7 & 69.8 & 64.7 & 80.2 & \textbf{91.3} \\ 
    \bottomrule
    \end{tabular}
\end{table}

Across all data ratios, SVM achieves the best performance, and the relative ranking between models closely mirrors their LogME scores, validating LogME as a reliable predictor of downstream performance. Additionally, accuracy saturates beyond approximately 5\% labeled data (e.g., Mantis stabilizes near 91\% from 10\%–80\% data), indicating that the pre-trained features already provide sufficient discriminative information, with limited gains from additional samples.

\subsection{Foundation Model Fine-Tuning Results}
To evaluate the effectiveness of the proposed foundation model, a comparative study is conducted against conventional deep learning architectures (CNN, LSTM, and Transformer~\cite{vaswani2017attention}) and recent state-of-the-art time-series models (SOTA), including TSSequencerPlus~\cite{tatsunami2022sequencer}, MINIROCKET~\cite{tan2022multirocket}, TSPerceiver~\cite{jaegle2021perceiver}, XCM~\cite{fauvel2021xcm} and PatchTST~\cite{Yuqietal-2023-PatchTST}. These SOTA approaches maintain the general design principles of deep learning while introducing task-oriented architectural refinements to enhance time-series representation capability.

\subsubsection{Fine-Tuning MOMENT}

\begin{figure}[!t]
\centering
\includegraphics[width=1.0\columnwidth]{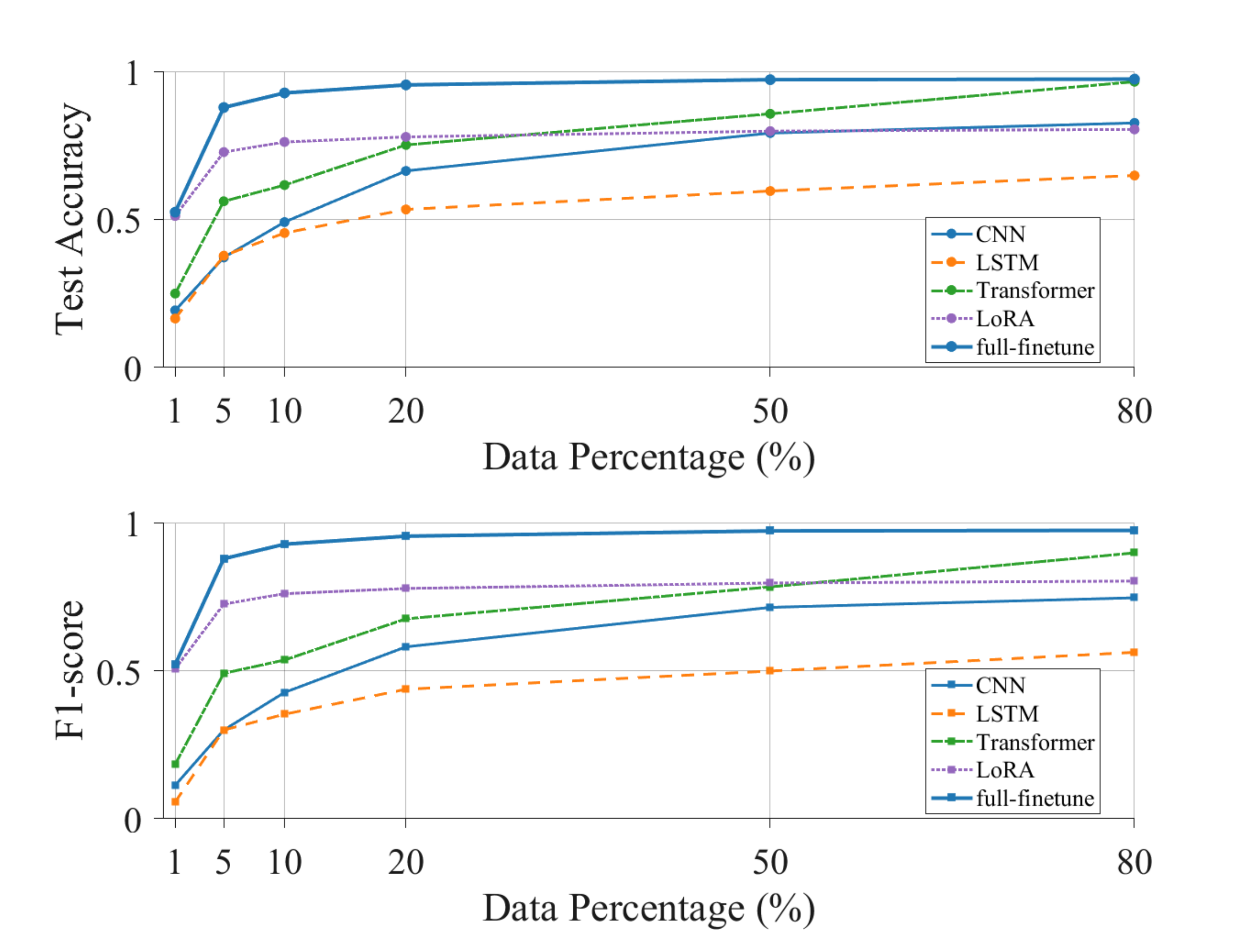}
\caption{Performance comparison between the fine-tuned MOMENT and traditional deep learning models across varying training data ratios.}
\label{fig: compare}
\vspace{-0.2cm}
\end{figure}

As confirmed by the LogME assessment in the previous subsection, the Mantis model generally outperforms MOMENT in classification-oriented pre-training. Nevertheless, the MOMENT model exhibits clear advantages over conventional deep learning baselines. Fig.~\ref{fig: compare} presents classification accuracy under varying training data ratios. The “Transformer” baseline is implemented using the MOMENT architecture without pre-training, isolating the contribution of pre-trained parameters.

For each experiment, 20\% of the dataset is reserved for testing, while the remaining 80\% is split into training (80\%) and validation (20\%) subsets. Different proportions of the training data are utilized to simulate limited-data conditions. The fine-tuned MOMENT foundation model demonstrates significant performance gains, particularly when the available training data are scarce (1\%–5\% of total samples). This enhancement stems from the large-scale pre-training stage, which provides transferable feature representations that improve data efficiency and convergence stability.

In addition, parameter-efficient fine-tuning using LoRA achieves accuracy comparable to full fine-tuning under small-sample conditions, verifying its suitability for resource-constrained deployment. As the training data scale increases, the baseline Transformer without pre-training gradually narrows the gap but consistently trails the fine-tuned foundation model. These results confirm that pre-training plays a pivotal role in achieving robust generalization across data regimes.

\subsubsection{Fine-tuning Mantis}

\begin{figure*}[!t]
\centering
\includegraphics[width=1.9\columnwidth]{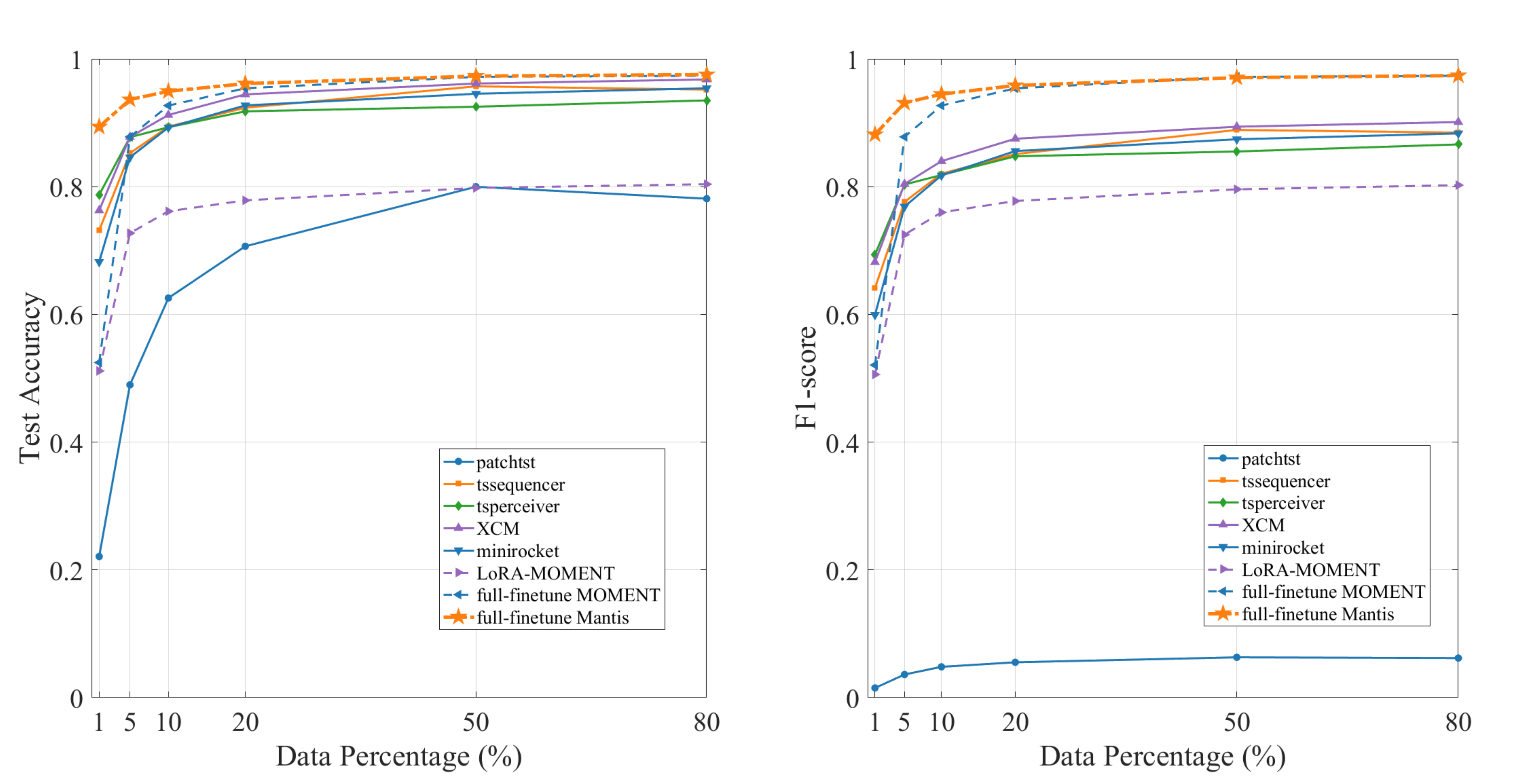}
\caption{Comparison between time series foundation models (MOMENT and Mantis) and state-of-the-art models (TSSequencerPlus, MINIROCKET, TSPerceiver, XCM and PatchTST) across different training data ratios.}
\label{fig:m5}
\vspace{-0.2cm}
\end{figure*}

The MOMENT model, pre-trained via masked reconstruction, demonstrates strong capability in prediction and interpolation tasks. In contrast, the Mantis model adopts a contrastive pre-training strategy that inherently yields more discriminative representations for classification. Fig.~\ref{fig:m5} compares the fine-tuned Mantis model with recent state-of-the-art (SOTA) architectures.

Across all data ratios, Mantis consistently achieves the highest classification accuracy. Even under zero-shot evaluation, it surpasses all SOTA baselines trained with limited samples, highlighting its strong generalization under data-scarce conditions. Although the fully fine-tuned MOMENT model ranks second, it requires larger datasets to fully exploit its reconstruction-based representations. When data are insufficient, MOMENT lags behind contrastive models optimized for classification. Additionally, PatchTST records the lowest F1 score because its original design targets forecasting rather than discriminative tasks; substituting its prediction head with a classification layer inevitably degrades performance.

Notably, the Mantis model contains only 8M parameters—significantly fewer than MOMENT—making it highly efficient for fine-tuning and deployment. The model converges rapidly and can be trained directly on a CPU without the need for parameter-efficient methods such as LoRA. Full-parameter fine-tuning thus remains both feasible and effective. These results confirm that an appropriately pre-trained foundation model can achieve high diagnostic accuracy with minimal computational resources, offering a practical solution for real-time fault classification in industrial environments.

\subsubsection{Scaling Analysis}
\begin{figure}[!t] 
\centering 
\includegraphics[width=0.9\columnwidth]{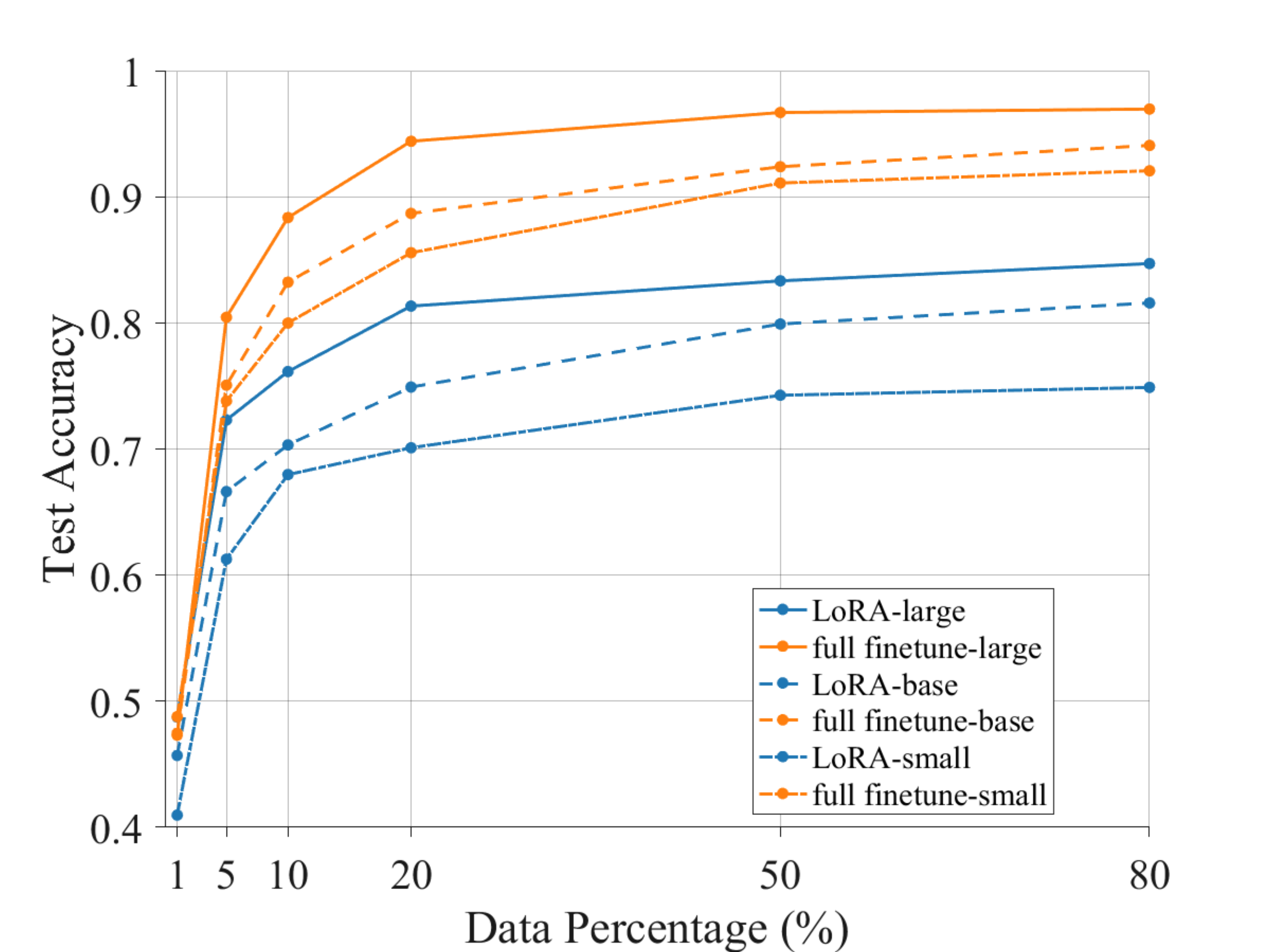} 
\caption{Performance comparison of MOMENT models with different parameter sizes: small (35.42M), base (109.77M), and large (341.42M).} \label{fig:m4} 
\vspace{-0.2cm} 
\end{figure}

A key advantage of the proposed foundation model is its scalability with respect to model size and computational budget. The relationship among computational cost ($C$), model parameters ($N$), and dataset size ($D$) follows the OpenAI scaling law for decoder-only models~\cite{kaplan2020scaling}:

\begin{equation}
C \approx 6ND
\end{equation}
where $C$ denotes the total cost of training before the training, $N$ is the parameter count, and $D$ is the volume of processed data (batch size $B$ multiplied by the gradient steps $S$). Model performance also follows a complementary scaling form:

\begin{equation}
L(x) = L_{\infty} + \left( \frac{x_0}{x} \right)^{\alpha}
\end{equation}
where $L_{\infty}$ represents the irreducible loss due to data noise, and the second term captures the reducible loss that decreases as computational scale grows. Increasing $N$ or $D$ therefore reduces training loss and improves accuracy.

Fig.~\ref{fig:m4} illustrates how these trends manifest in practice. Across all data ratios, the large model consistently achieves the best performance, followed by the base and small variants. Accuracy rises sharply between 1\% and 10\% data and then begins to saturate, reflecting the expected diminishing returns as models approach their scaling limit.

The figure further highlights the effect of parameter-efficient fine-tuning. For each model size, the LoRA curves closely match their full fine-tuning counterparts, with only minor deviations even in low-data regimes. In particular, the large model under LoRA reaches over 70\% accuracy at 5\% data—nearly identical to full fine-tuning—while requiring far fewer trainable parameters. This demonstrates that LoRA preserves most of the performance benefits of full tuning while substantially reducing memory and compute overhead, making it especially suitable for industrial deployments with limited resources.

\section{Conclusion}\label{cons}
This work applies time series foundation models to motor condition monitoring and establishes a unified workflow integrating model assessment and supervised fine-tuning. The proposed approach achieves high diagnostic accuracy using only about 1\% of the available training data, effectively addressing the challenges of limited fault samples and class imbalance. Furthermore, it supports direct fine-tuning on standard CPUs, enabling faster adaptation compared with deep learning models trained from scratch. The LogME model assessment strategy efficiently identifies suitable pre-trained models, eliminating the need for iterative trial-and-error. Experimental validation confirms the strong generalization and scalability of the foundation model, offering valuable insights for developing efficient, adaptable, and data-efficient solutions for intelligent motor condition monitoring in industrial applications.

\bibliographystyle{ieeetr}
\bibliography{ref}

\vfill

\end{document}